\documentclass[twocolumn,aps,prl,reprint,superscriptaddress]{revtex4-2}

\usepackage[utf8]{inputenc} 
\usepackage[T1]{fontenc}      
\usepackage{epsf,graphicx}
\usepackage{natbib}
\usepackage{multirow}
\usepackage{amsmath,gensymb,amssymb}
\usepackage{siunitx}
\usepackage[bottom]{footmisc}
\usepackage{lipsum}
\usepackage{dcolumn}
\usepackage{xcolor}
\usepackage{upgreek}
\usepackage{float,ulem}

\newcommand{\orcid}[1]{%
  \href{https://orcid.org/#1}{\raisebox{-0.05em}{\includegraphics[height=1em]{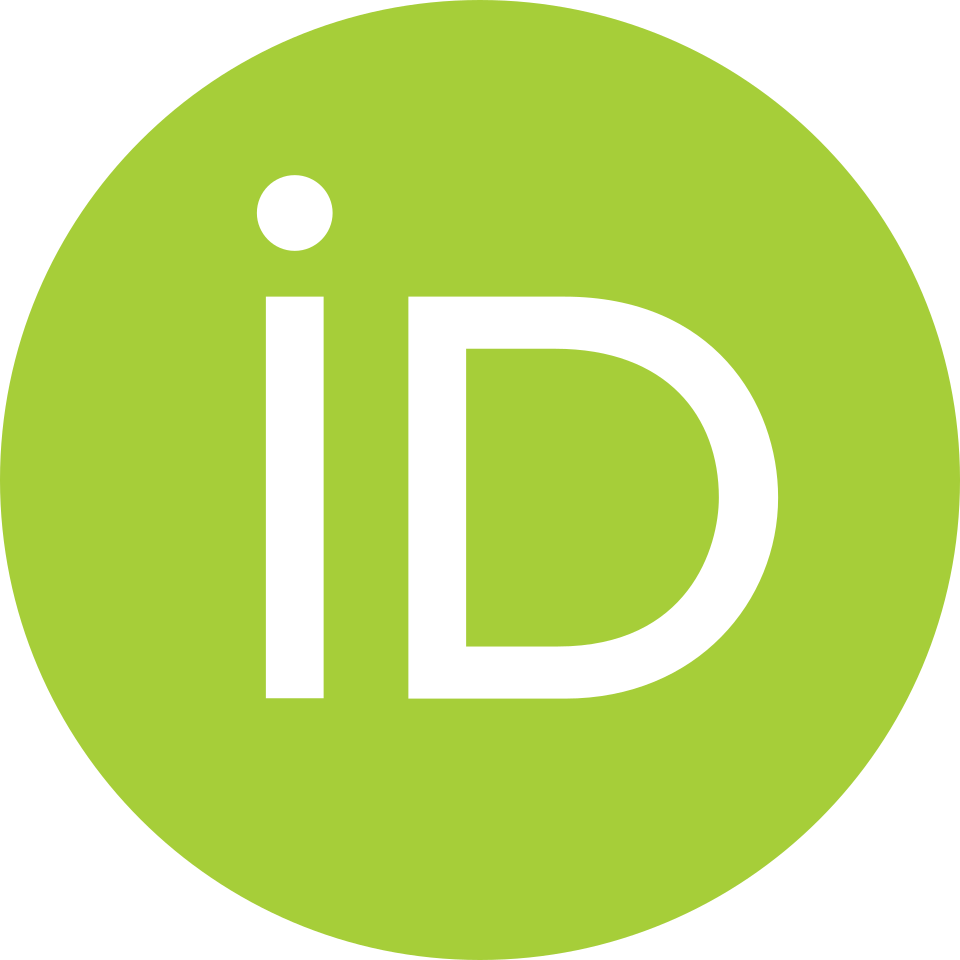}}}%
}

\usepackage{bm}
\usepackage[colorlinks=true,allcolors=blue]{hyperref}

\usepackage{natbib}
\usepackage{color}
\usepackage[type={CC},modifier={by-nc-sa},version={4.0},]{doclicense}
\definecolor{linkcolor}{rgb}{0,0,0.6}

\begin{document}
\title{Experimental evidence of Kelvin wave turbulence along a vortex core}

\author{Jason Barckicke\orcid{0009-0001-5935-3498}}
\affiliation{Universit\'e Paris Cit\'e, CNRS, MSC Laboratory, UMR 7057, F-75013 Paris, France}
\affiliation{LPENS, ENS, Universit\'e PSL, CNRS, Paris, France}

\author{Christophe Gissinger\orcid{0000-0002-1734-6716}}
\affiliation{LPENS, ENS, Universit\'e PSL, CNRS, Paris, France}
\affiliation{Institut Universitaire de France}

\author{Eric Falcon\orcid{0000-0001-9640-9895}}
\email{eric.falcon@u-paris.fr}
\affiliation{Universit\'e Paris Cit\'e, CNRS, MSC Laboratory, UMR 7057, F-75013 Paris, France}


\begin{abstract}
Wave turbulence is a regime of interacting nonlinear waves occurring in most physical systems. Kelvin waves are helical distortions that propagate along vortex filaments and are believed to play a central role in quantum turbulence up to atmospheric vortices. Yet, Kelvin wave turbulence has remained inaccessible to direct experimental observation. Here, we report the first direct experimental observation of Kelvin-wave turbulence along a single vortex filament in a classical fluid under controlled conditions. Using high-resolution spatiotemporal measurements, we resolve Kelvin-wave dynamics over a broad range of scales and obtain wave-amplitude spectra consistent with the predicted weak-turbulence cascade. We identify six-wave resonant interactions as the mechanism driving this energy transfer, providing direct experimental support for a long-standing prediction of weak-turbulence theory. These results establish an experimental platform for investigating energy transport along vortex filaments, with broader implications for both classical and quantum turbulent systems.
\end{abstract}

\maketitle

\textit{Introduction---}
Kelvin waves are helical perturbations propagating along vortex lines. They were first introduced in 1880 by Lord Kelvin in the context of classical vortex dynamics~\cite{Thomson1880} and later recognized as a key ingredient of quantum turbulence in superfluids, where vorticity is confined to quantized vortex lines~\cite{DonnellyBook1991,Kivotides2001,VinenJLTP02,BarenghiBook2023}. At scales smaller than the intervortex distance, vortex reconnections generate Kelvin waves, and their nonlinear interactions are expected to drive a turbulent energy cascade along individual filaments, commonly referred to as Kelvin-wave turbulence~\cite{SvistunovPRB1995,KozikPRL04,BoffettaJLTP2009,LvovLTP2010}. This process is believed to mediate the transfer of energy toward small scales in quantum turbulence, providing the route to dissipation where the energy is ultimately radiated as sound (phonon emission)~\cite{Kivotides2001,VinenJLTP02,BarenghiBook2023}.

In superfluids, weak turbulence theory predicts a Kelvin-wave turbulence regime characterized by self-similar energy spectra arising from resonant nonlinear wave interactions~\cite{NazarenkoBook,ZakharovBook}. Depending on the dominant interaction order, distinct stationary cascades have been proposed: six-wave interactions leading to a wave spectrum $E_{KS}\propto k^{-7/5}$~\cite{KozikPRL04,BoffettaJLTP2009}, and effective four-wave processes yielding $E_{LN}\propto k^{-5/3}$~\cite{LvovLTP2010,LauriePRB10}. Numerical studies support the latter scaling within the Gross-Pitaevskii framework for single~\cite{KrstulovicPRE12} and vortex-tangle configurations~\cite{ClarkPRA15,VilloisPRE16,MullerPRB20}, as well as in Biot-Savart vortex simulations~\cite{BouePRB11,BaggaleyPRL12,BaggaleyPRB14,KondaurovaPRB2014}. Spectrum-prefactor scalings have also been obtained numerically~\cite{BouePRB11,BaggaleyPRB14,MullerPRB20}. Despite these advances, experimental evidence of Kelvin-wave turbulence is still lacking. Kelvin waves are difficult to access on quantum vortices because of their nanometric core size~\cite{Fonda2014,Peretti2023,Minowa2025}. By contrast, the Kelvin-wave dispersion relation along the macroscopic core of a classical vortex has recently been measured experimentally~\cite{BarckickeNaturePhys2026}, opening access to the nonlinear regime.

Here, we report the first experimental evidence of Kelvin-wave turbulence developing along the core of a single classical vortex. By directly exciting and measuring nonlinear Kelvin-wave dynamics, we observe the emergence of a broadband power-law spectrum consistent with theoretical predictions for a weak wave-turbulence cascade. Because Kelvin-wave dynamics is intrinsic to a vortex filament, our results capture the universal mechanisms underlying filament excitations and thus lend experimental support to the Kelvin-wave cascade scenario proposed for quantum turbulence.

\textit{Theoretical predictions---}
For an inviscid, incompressible fluid, let us assume a Rankine vortex: a solid-body rotation inside the core ($r<a_0$)  and irrotational flow outside. For infinitesimal velocity (pressure) perturbations $u_i(r)e^{i(kz + m\theta + \omega t)}$ to the Euler equations, in cylindrical coordinates ($r$, $\theta$, $z$), with continuity of velocity (pressure) at the perturbed surface, the dispersion relation $\omega(k)$ of linear Kelvin waves reads~\cite{Thomson1880,Saffman1993}
\begin{equation}
\frac{(\omega+m\Omega_0)^2}{4\Omega_0^2-\widetilde{\omega}^2}\left[\frac{\beta a_0 J'_{|m|}[\beta a_0]}{J_{|m|}[\beta a_0]} + \frac{2\Omega_0 m}{\widetilde{\omega}} \right] = -x\frac{K'_{|m|}[x]}{K_{|m|}[x]} \ .
\label{fullDispRel}
\end{equation}
$\Omega_0\equiv \Gamma/(2\pi a_0^2)$ is the angular rotation at core radius $a_0$, $\Gamma$ is the flow circulation (the analog of the quantum of circulation of a quantum vortex in superfluids), $x \equiv a_0|k|$, $\widetilde{\omega}\equiv \omega+m\Omega_0-kv_z$, $k$ is the wave number, $\omega$ the angular frequency, $v_z$ is a downward (positive) fluid velocity inside the core, $\beta \equiv k\sqrt{4\Omega_0^2/\widetilde{\omega}^2-1}$, and $J_m$ (resp. $K_m$) is the Bessel (modified Bessel) function of the first (second) kind of order $m$. 
Bending waves correspond to the $m=1$ mode and feature a helical core displacement at constant radius. Such waves are predicted to counterrotate with respect to the flow, as observed recently~\cite{BarckickeNaturePhys2026}. Modes $m\neq 1$ correspond to core radius fluctuations. For $m=1$, in the long wavelength limit $(ka_0 \ll 1$), Eq.~\eqref{fullDispRel} reduces to $\omega=- \frac{\Gamma}{4\pi}k^2\left[\ln\left(\frac{2}{ka_0}\right)-\gamma +1/4\right]$ with $\gamma\simeq 0.577$ the Euler constant, and in the small-scale limit ($ka_0 \gg 1$), to $\omega\simeq kv_z+\Omega_0$.

In 1D, $N$-wave resonant nonlinear interactions satisfy
\begin{equation}
\omega_1 \pm \omega_2 \pm \cdots \pm \omega_{\it N} = 0  \ \ \mathrm{and} \ \ k_1 \pm k_2 \pm \cdots \pm k_{\it N} = 0 
\label{Nwave}
\end{equation}
where $\omega_i\equiv \omega(k_i)$, $N\geq 3$. For weakly nonlinear waves, and $\omega(k)$ as in  Eq.~\eqref{fullDispRel} with $ka_0 \ll 1$, $m=1$, weak turbulence theory gives the energy spectrum $E(k)$ of Kelvin wave turbulence on a 1D vortex line~\cite{KozikPRL04,NazarenkoBook,LvovLTP2010}. 
A six-wave local resonant interaction process was first derived by Kozik-Svistunov (KS) for Kelvin wave turbulence~\cite{KozikPRL04}, i.e., for the wave-amplitude power spectrum,
\begin{equation}
S^{KS}_{\eta}(k)\sim k^{-17/5} \ \ \mathrm{and}\ \  S^{KS}_{\eta}(\omega) \sim \omega^{-11/5} \ . 
\label{KSspectrum}
\end{equation} 
L'vov and Nazarenko (LN) later pointed out the infrared divergence of such a model and derived a nonlocal six-wave interaction (equivalently, an effective local four-wave process), leading to the Kelvin-wave turbulence spectrum~\cite{NazarenkoBook,LvovLTP2010}
\begin{equation}
S^{LN}_{\eta}(k)\sim\ k^{-11/3} \ \ \mathrm{and}\ \ S^{LN}_{\eta}(\omega)\sim\omega^{-7/3} \ . 
\label{LNspectrum}
\end{equation}
 
However, to date, no experiment has observed this Kelvin wave turbulence. 
Besides, when nonlinear interactions are strong enough, a critical-balance regime is predicted to occur with an energy spectrum $E^{cb}\propto k^{-1}$, i.e.,  $S^{cb}_{\eta}\propto k^{-3}$~\cite{VinenPRL03}.

\begin{figure}[t!]
     \includegraphics[width=0.85\linewidth]{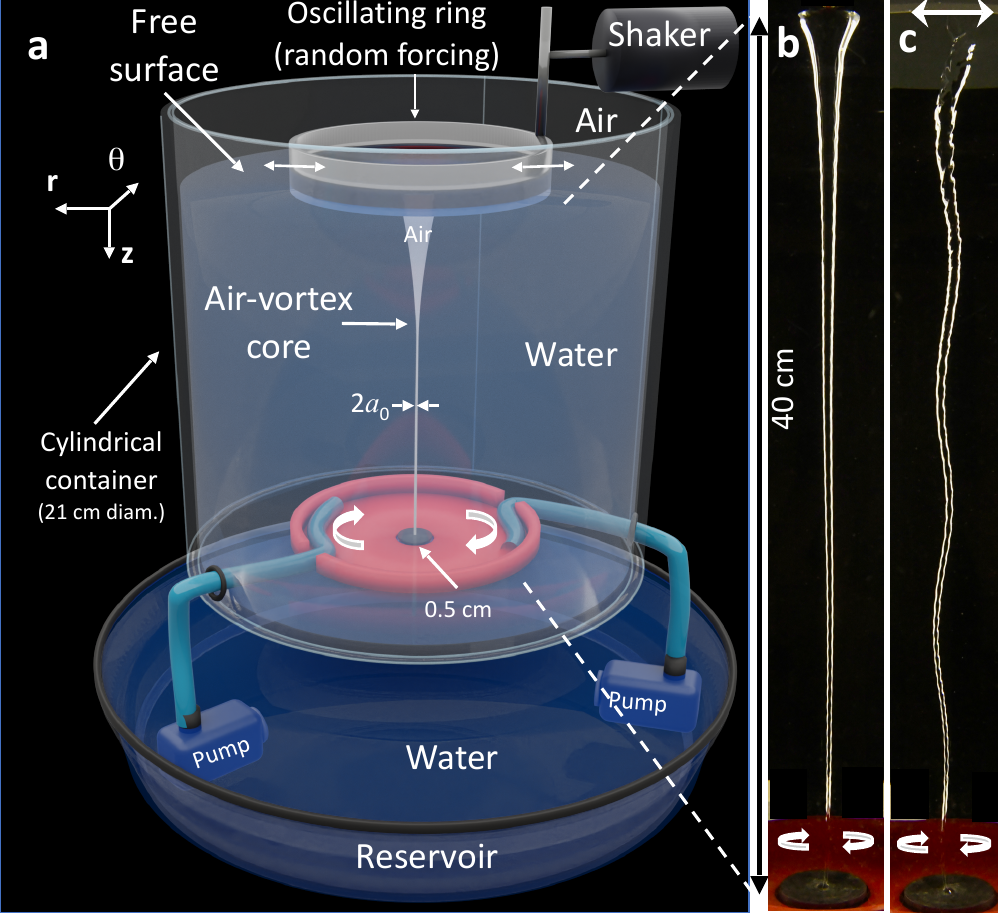}
  \caption{(a) Schematic of the experimental setup to observe Kelvin wave turbulence. (b) No forcing: Steady air-vortex core. (c) With forcing: cascade of bending Kelvin waves.} 
    \label{setup}   
\end{figure}

\begin{figure}[t!]
     \includegraphics[width=1\linewidth]{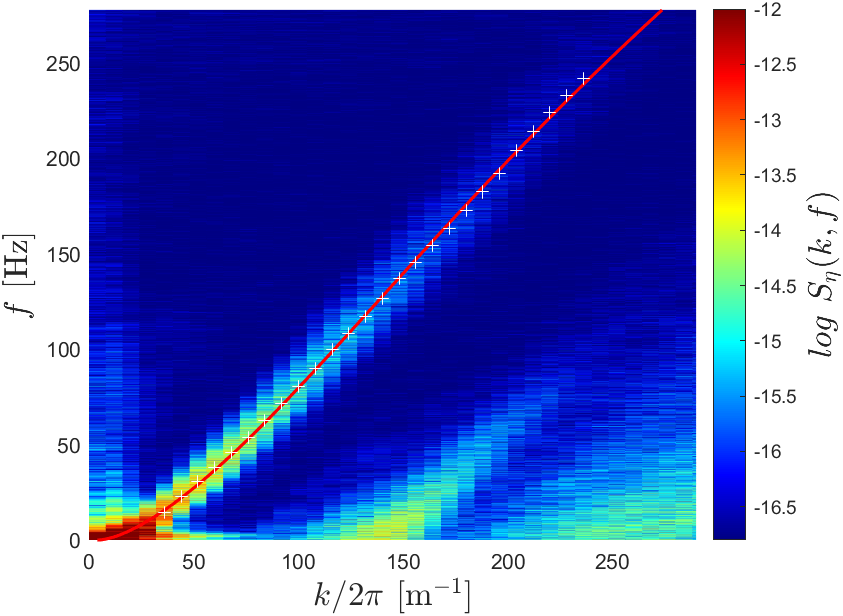}
  \caption{Spatiotemporal power spectrum of the vortex displacement $S_\eta(k,\omega)$ showing the dispersion relation of Kelvin waves. $\omega=2\pi f$. Random forcing frequency $f_p\in [1,3]$~Hz and amplitude $\sigma_{\eta}=1.8$~mm. White crosses: Maxima of Gaussian fits along the experimental $m=1$ branch. Other patches correspond to higher $m$ modes. Red line: theoretical dispersion relation of Eq.~\eqref{fullDispRel} with measured parameters $a_0=1.47$~mm, $\Gamma=0.018$~m$^2$/s, and $\mathrm{v}_z=-0.63$~m/s.}  
    \label{disprelation} 
\end{figure}

\textit{Experimental setup---} 
The experiment is designed to generate a stable, elongated vortex filament enabling direct measurements of Kelvin-wave dynamics over a broad range of scales. A stationary bathtub-type vortex is produced using water within a cylindrical Plexiglas tank of diameter 21 cm and height 45 cm [see Fig.~\ref{setup}(a)], embedded in a larger rectangular reservoir ($50 \times 50 \times 50$~cm$^3$ - not shown). Water is injected at a controlled rate through four inlets symmetrically arranged around a circular drain (5 mm in diameter) at the bottom of the tank, using two independent pumps. This configuration allows precise control of the vortex circulation $\Gamma$ (through the pump flow rate) while avoiding the solid-body rotation characteristic of conventional rotating-tank experiments, which tends to suppress Kelvin waves. The flow converges toward the drain and forms a thin, steady vortex with a central air core, whose height reaches $H=40$~cm, two orders of magnitude larger than its core radius [see Fig.~\ref{setup}(b)]. Due to the combined hydrostatic pressure, rotational flow, and localized suction at the drain, the air-core radius $a(z)$ increases slightly with altitude $z$, from approximately 1 mm near the drain to a few millimetres in the central region of the tank. Its mean value is $a_0\equiv \langle a\rangle_z$. The circulation $\Gamma$ is fixed and measured to be $\Gamma= 1.8\ 10^{-2}$~m$^{2}$~s$^{-1}$ (see End Matter). The angular rotation is $\Omega_0\approx 10^3$~s$^{-1}$. The nondimensional parameters are $\omega/\Omega_0\in [0,1.5]$ and $|k|a_0\in [0,4]$.  

Bending Kelvin waves are excited by applying low-frequency random horizontal oscillations ($f_p$) at the top of the vortex using an immersed annulus connected to an electromechanical shaker [see Fig.~\ref{setup}(a)]. The random forcing frequency and amplitude are $f_p\in[1,3]$~Hz and $\sigma_{\eta}\in [0.9,1.9]$~mm, respectively. This forcing generates bending waves propagating downward along the vortex filament [see Fig.~\ref{setup}(c)].  
The vortex shape is imaged through the air-water interface using an optical camera (Basler 480~fps, 3~Mpix) positioned perpendicular to the vortex axis, at a height of 18 cm.  The sharp optical contrast at the interface allows time-resolved tracking of the vortex-core horizontal displacement $\eta(z, t)$ over a vertical window of length $L=20.6$~cm, where $\eta(z, t)=[r_i(z,t,\theta=0)-r_i(z,t,\theta=\pi)]/2$ and $r_i$ is the radial interface position. Spatiotemporal Fourier analysis of these quantities provides direct access to the Kelvin-wave energy cascade (see movie in Supp. Mat~\cite{SuppMat}) and to the wave interactions (see below).  Acquisitions last $\mathcal{T}=1$~min for the wave spectrum, 15~min for its probability distribution, or 75 min to converge high-order correlation estimators. The wave steepness is weak enough ($k_p\sigma_{\eta}\approx 0.1$ with $\sigma_{\eta}\equiv\sqrt{\langle \eta^2 \rangle_{t,z}}\simeq 2$~mm) to ensure that waves remain weakly nonlinear.  

\begin{figure}[t!]
\includegraphics[width=1\columnwidth]{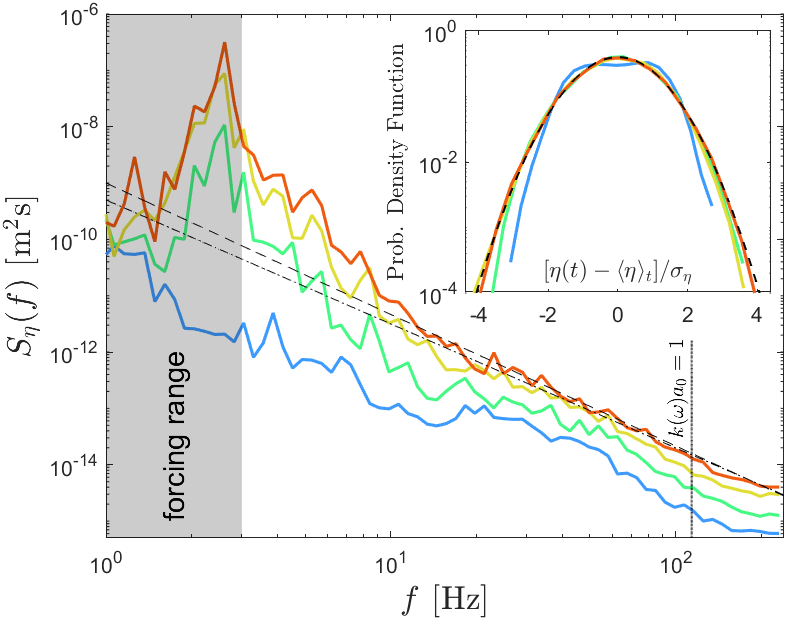}
\caption{\label{Sf}Frequency power spectrum of the wave amplitude $S_\eta(f)$ for different wave-forcing strengths $\sigma_{\eta}=0.9$, 1.3, 1.4, and 1.9 mm (from bottom to top). Fixed $\Gamma$ and $a_0=1.3$~mm. Gray region: random forcing range $f_p\in [1,3]$~Hz. The dashed line has a slope of $-7/3$ as in Eq.~\eqref{LNspectrum} and the dash-dotted line, a slope of $-11/5$ as in Eq.~\eqref{KSspectrum}.  Inset: Probability density functions (PDF) of Kelvin wave amplitudes. Same forcing.}  
\end{figure}

\textit{Energy cascade---}We first compute the 2D-Fourier transform of the vortex-core displacement $\eta(z,t)$ to get the spatiotemporal power spectrum $S_{\eta}(k,\omega)\equiv |\eta(\omega,k)|^2/(L\mathcal{T})$ of Kelvin waves, as shown in Fig.~\ref{disprelation}. The Kelvin wave energy is observed to spread from the forcing scales ($f_p\in[1,3]$~Hz) to smaller scales, over nearly two decades in frequency and wavenumber, along the theoretical dispersion relation of Eq.~\eqref{fullDispRel} (red line), indicating a wave turbulence cascade. Such a prediction is quadratic in the small-$k$ limit and roughly linear for $ka_0\sim 1$, in agreement with experimental data. 
Other visible energy patches correspond to higher azimuthal modes, such as double helix ($m=2$ flattening modes)~\cite{BarckickeNaturePhys2026}.  

\begin{figure}[t!]
\includegraphics[width=1\columnwidth]{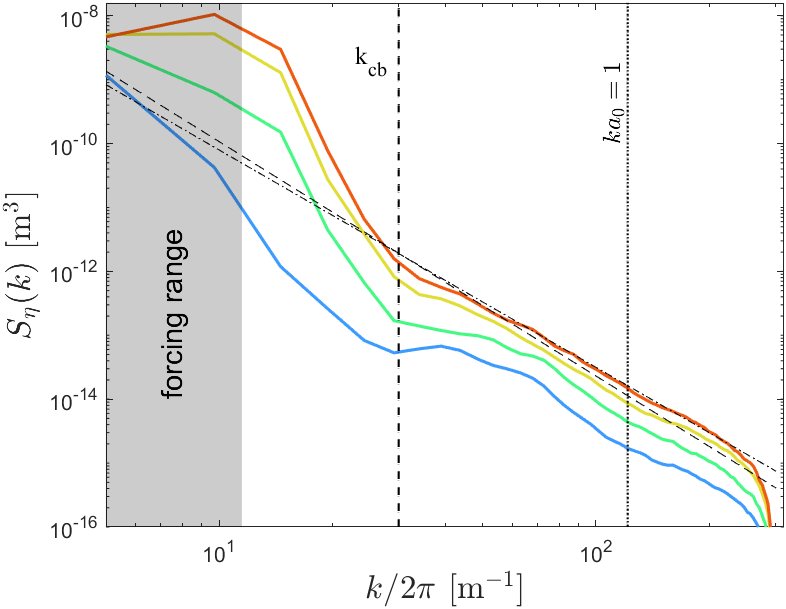} 
\caption{\label{Sk}Spatial power spectrum of the wave amplitude $S_\eta(k)$ for the same wave-forcing strengths as in Fig.~\ref{Sf}. Fixed $\Gamma$ and $a_0=1.3$~mm. Gray region indicates the corresponding forcing range $k_p/(2\pi)\in [3,11]$~m$^{-1}$. The dashed line has a slope of $-11/3$ as in Eq.~\eqref{LNspectrum} and the dash-dotted line, a slope of $-17/5$ as in Eq.~\eqref{KSspectrum}. The dotted vertical line indicates $ka_0=1$.}
\end{figure}

Figure~\ref{Sf} shows the frequency power spectrum $S_{\eta}(\omega)$ of Kelvin waves computed by integrating the spatiotemporal spectrum over $k$ as $S_{\eta}(\omega) \equiv \int S_{\eta}(k,\omega)dk$. For strong enough forcing (gray region), the wave spectrum displays a frequency power-law cascade over one decade in frequency, in very good agreement with theoretical predictions $S_{\eta}^{LN}(\omega)\propto \omega^{-7/3}$ of Eq.~\eqref{LNspectrum} (dashed line) or $S_{\eta}^{KS}(\omega)\propto \omega^{-11/5}$ of Eq.~\eqref{KSspectrum} (dash-dotted line). This confirms the observation of Kelvin wave turbulence. For any propagating waves, frequency and spatial spectra are related through the dispersion relationship $\omega \propto k^{\zeta}$. Specifically, if $S_\eta(\omega) \propto \omega^\alpha$ and $S_\eta(k) \propto k^\beta$, dimensional analysis then leads to $\beta = \zeta(\alpha + 1)-1$~\cite{ZakharovBook}. Thus, for Kelvin waves ($\omega \propto k^{2}$, i.e., $\zeta=2$) and for $\alpha = -7/3$, one should have $\beta = -11/3$ as weak Kelvin wave turbulence theory indeed predicts~\cite{LvovLTP2010}. To verify this consistency, we experimentally access the spatial power spectrum $S_{\eta}(k)$ by integrating the spatiotemporal spectrum over frequency as $S_{\eta}(k) \equiv \int S_{\eta}(k,\omega)d\omega$. Figure~\ref{Sk} shows the spatial spectrum for different forcing strengths. For strong enough forcing, the wave spectrum displays a wavenumber power-law scaling over one decade in $k$, in good agreement with the theoretical predictions $S_{\eta}^{LN}(k)\propto k^{-11/3}$ of Eq.~\eqref{LNspectrum} (dashed line) and $S_{\eta}^{KS}(k)\propto k^{-17/5}$ of Eq.~\eqref{KSspectrum} (dash-dotted line). Typically, the cascade extends from a wavelength approximately three times the vortex core size to one-third of that size. The low-$k$ spectrum part ($\lesssim 15$~m$^{-1}$) is reminiscent of the rotary sloshing resonance at $f_0\approx2.5$~Hz within the forcing frequency band. Note that the theoretical spectrum exponents in Eqs.~\eqref{LNspectrum} and \eqref{KSspectrum} are too close ($\approx-3.7$ and $-3.4$ in $k$ and $-2.3$ and $-2.2$ in $\omega$) to be experimentally distinguishable. 

\begin{figure}[t!]
\includegraphics[width=1\columnwidth]{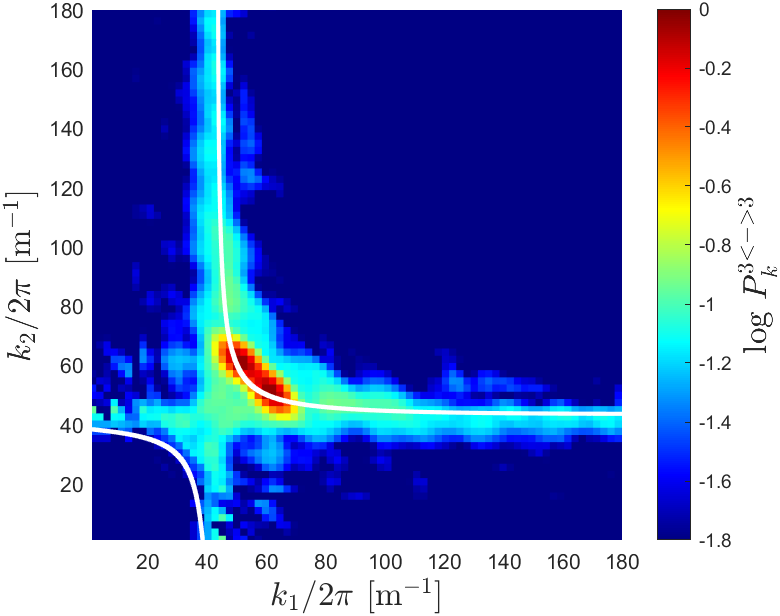} 
\caption{\label{Penta}Pentacoherence $P(k_1,k_2,k_3,k_4,k_5)$ highlighting six-wave interactions ($3\leftrightarrow 3$) for fixed $k_3=70$, $k_4=50$, and $k_5=60$~m$^{-1}$. $\sigma_{\eta}=1.4$~mm. $a_0=1.3$~mm. Solid lines: resonant interaction solutions of Eq.~\eqref{Nwave} with $N=6$ and $\omega(k)$ as Eq.~\eqref{fullDispRel}. Trivial wave interactions correspond to red points ($k_1=60$; $k_2=50$) and ($k_1=50$; $k_2=60$). Logscale colorbar.}
\end{figure}

\textit{Resonant wave interactions---} For 1D Kelvin waves with $\omega(k)$ as in Eq.~\eqref{fullDispRel}, the resonance conditions of Eq.~\eqref{Nwave} admit nontrivial solutions only for $N = 6$, as odd-order resonant interactions ($N = 3$, 5) are forbidden~\cite{NazarenkoBook}, while four-wave resonances are trivial. 
This six-wave process is hence expected to be the dominant one generating Kelvin wave turbulence. To verify the existence of six-wave interactions, we compute the normalized sixth-order correlation (or pentacoherence) in $k$ of the wave amplitudes. To our knowledge, this correlation has never been used before to analyze wave turbulence. Analogous to the bi- and tri-coherence used to explore lower-order correlations in other wave turbulence systems~\cite{FalconARFM2022,RicardEPL2021}, we define pentacoherence as
\begin{equation}
P=\frac{|\langle \eta_{k_1}^*\eta_{k_2}^*\eta_{k_3}^*\eta_{k_4}\eta_{k_5}\eta_{k_1+k_2+k_3-k_4-k_5}\rangle|}{\sqrt{\langle| \eta_{k_1}\eta_{k_2}\eta_{k_3}|^2\rangle\langle|\eta_{k_4}\eta_{k_5}\eta_{k_1+k_2+k_3-k_4-k_5}|^2\rangle}} {\rm \ ,}
 \end{equation}
where $\eta_k=\int \eta(z,t)e^{-ikz}\,dz$ is the spatial Fourier transform of wave amplitudes at time $t$, $^\ast$ denotes its complex conjugate, and $\langle \cdot \rangle$ represents an average over $\mathcal{T}=75$~min. The normalization is chosen to bound $P$ between 0 (no correlation) and 1 (perfect correlation). 
Figure~\ref{Penta} shows $P(k_1,k_2,k_3,k_4,k_5)$ for fixed $k_3$, $k_4$, and $k_5$ within the inertial range. Pentacoherence shows the occurrence of a large number of local six-wave resonances ($3\leftrightarrow 3$) well described by solutions of Eq.~\eqref{Nwave} (see white solid lines) with $N=6$ and $\omega(k)$ as Eq.~\eqref{fullDispRel}. Note that the vertical asymptote of the solution $k_1+k_2+k_3=k_4+k_5+k_6$ corresponds to nonlocal six-wave interactions ($k_2,k_6\gg k_1,k_3,k_4,k_5$) and thus to effective four-wave {\it quasiresonant} interactions $k_1+k_3 \simeq k_4+k_5$, as $k_2\simeq k_6$. Same for the horizontal asymptote with indices $1$ and $2$ swapped. This demonstrates that the mechanism driving Kelvin wave turbulence is six-wave resonant interactions, as predicted theoretically. We also experimentally verified that no four-wave {\it resonant} interaction occurs (see End Matter).

\begin{figure}[t!]
\includegraphics[width=1\columnwidth]{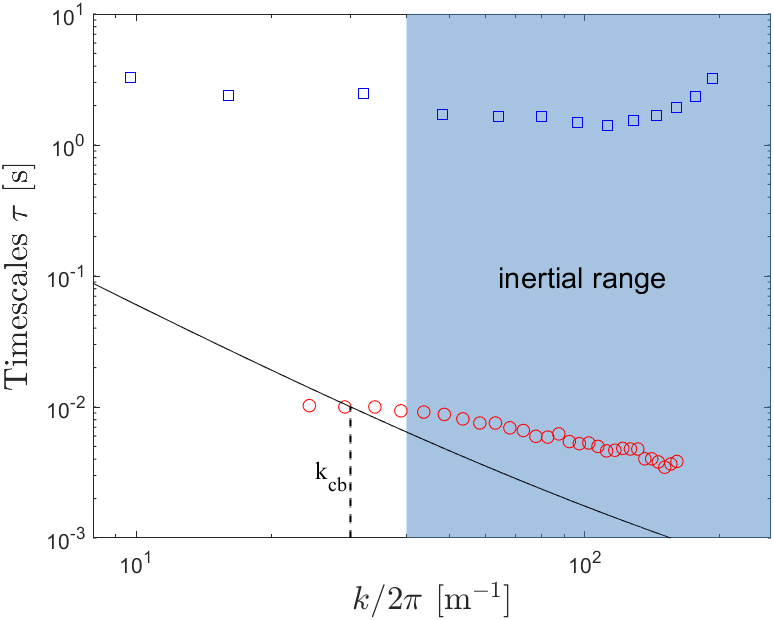}
\caption{\label{timescales}Wave-turbulence timescale separation. Black solid line: linear timescale $1/\omega$. Experimental nonlinear timescale $\tau_{\mathrm{nl}}$ (circles) and dissipation time $\tau_{\mathrm{diss}}$ (squares). Blue region: inertial range of Fig.~\ref{Sk}. A critical balance occurs for $\tau_{\mathrm{lin}}(k_{cb})=\tau_{\mathrm{nl}}(k_{cb})$.} 
\end{figure}

{\textit{Timescales---}Finally, we verify the validity of the main weak-turbulence hypotheses in our experiment. First, the weakly nonlinear-wave assumption is verified, as wave steepness is $k_p\sigma_{\eta} \simeq 0.1$. Second, the probability distribution of Kelvin waves is well Gaussian, for strong enough forcing (see inset of Fig.~\ref{Sf}). Weak turbulence theory also assumes a timescale separation ($\forall k$ in the inertial range)~\cite{NazarenkoBook}, between the linear time $\tau_{\mathrm{lin}}$, the nonlinear time $\tau_{\mathrm{nl}}$, and the dissipation time $\tau_{\mathrm{diss}}$, as~\cite{FalconARFM2022,RicardPRF2023,SinglaPRF2026}
\begin{equation}
    \tau_{\mathrm{lin}}(k)\ll\tau_{\mathrm{nl}}(k)\ll\tau_{\mathrm{diss}}(k). 
    \label{scale}
\end{equation}
The nonlinear evolution is thus assumed to be slow relative to the fast linear oscillations (wave period) but short relative to the typical wave dissipation time, enabling an energy cascade in the inertial range. These timescales are plotted versus $k$ in Fig.~\ref{timescales}. 
The nonlinear time $\tau_{\rm nl}$ (circles) is experimentally inferred from the wave-energy broadening around the dispersion relation in Fig.~\ref{disprelation}, as $\tau_{\rm nl}=1/\delta\omega$, where $\delta\omega$ is the full-width-at-half-maximum of a Gaussian fit at each $k$. It always remains much smaller than the dissipation time $\tau_{\rm diss}$ (squares) over the inertial range (see Fig.~\ref{timescales}). The latter is inferred from decaying experiments and is found to be roughly independent of $k$, and of the same order of magnitude as the turbulent dissipation time $\tau_{\rm diss}=H^{1/3}k_0^{-2/3}\sigma_U^{-1}\simeq 1.5$~s, with $\sigma_U=\sigma_{\eta}\omega_0$ the velocity fluctuations and $k_0=k(\omega_0)$ the resonant mode~\cite{Gutierrez2016}. Such $k$-independent dissipation has also been observed in other wave-turbulence systems and attributed to finite-size effects~\cite{CazaubielPRL2019,RicardEPL2021}. We then compare $\tau_{\rm  nl}$ with the linear time $\tau_{\rm  lin}=1/\omega$ (black solid line). Figure~\ref{timescales} thus shows clearly that the timescale separation of Eq.~(\ref{scale}) is verified, within the inertial range of Fig.~\ref{Sk}. At larger scales, a critical balance occurs when $\tau_{\rm lin}(k_{\rm cb})=\tau_{\rm nl}(k_{\rm cb})$, thereby breaking the weak wave-turbulence cascade for $k<k_{\rm cb}$ (see Fig.~\ref{Sk}).

{\textit{Conclusions---}We reported the first direct experimental observation of Kelvin-wave turbulence along a vortex line. The measured wave spectra resolve nonlinear Kelvin-wave dynamics over a broad range of scales and display clear signatures of an energy cascade consistent with weak turbulence theory. The identification of six-wave resonant interactions as the mechanism driving this cascade provides strong experimental evidence for Kelvin-wave turbulence, while the main theoretical assumptions (timescale separation and weakly nonlinear waves with normal statistics) are validated quantitatively. These results thus elucidate the fundamental process governing energy transfer in nonlinear wave dynamics along classical vortices and, by extension, along superfluid vortices. 
Future experiments could investigate the existence of an inverse cascade regime~\cite{NazarenkoBook,ZhuPRL2023}, and, at strong forcing, of a critical balance state~\cite{ZhuRPP2026}. More broadly, this work will enable controlled investigations of solitons propagating along vortex filaments~\cite{HashimotoJFM1972,HopfingerJFM1982,SterkersArXiv}, and of linear and nonlinear interactions in a vortex array mediated by collective modes (Tkachenko waves), a phenomenon occurring in various domains involving vortex array dynamics, from classical fluids to superfluids and Bose-Einstein condensates~\cite{Sonin2013,CoddingtonPRL2003}.

\begin{acknowledgments}
\textit{Acknowledgments---}We thank G. Ricard for fruitful discussions, Y. Le Goas and A. Di Palma for their technical support. This work was supported by the Simons Foundation Project No. MPS-WT-00651463 (U.S.) on Wave Turbulence and the French National Research Agency (ANR Sogood Project No.~ANR-21-CE30-0061-04, ANR Lascaturb Project No.~ANR-23-CE30-0043-02, and ANR Provebact ANR-24-CE09-1394-02). C.G. acknowledges financial support from the Institut Universitaire de France.
\end{acknowledgments}

\textit{Data availability---}The data that support the findings of this article are not publicly available. The data are available from the authors upon reasonable request.

\section*{End Matter}  
\setcounter{figure}{0} 
\renewcommand{\thefigure}{EM\arabic{figure}} 
\textit{Theoretical backgrounds---}The energy spectrum $E(k)$ of Kelvin waves is related to the wave-amplitude power spectrum $S_{\eta}(k)$ by $E(k)=\Gamma \omega(k) S(k)$~\cite{KozikPRL04}, and $E(\omega)d\omega=E(k)dk$ using $\omega(k)$.

Note that the hollow-core vortex model (no fluid inside the core and irrotational flow outside) leads to a dispersion relation of linear Kelvin waves as~\cite{Thomson1880,Saffman1993,KrstulovicHDR2020}
\begin{equation}
\omega_{hc}=\frac{\Gamma}{2\pi a_0^2}\left[1-\sqrt{1+\frac{xK_{m-1}(x)}{K_{m}(x)}}\right] \ \ \ x\equiv a_0|k|
    \label{KWeq}
\end{equation}
which superimposes in Fig.~\ref{disprelation} for $a_0=0.7$~mm, $\Gamma=0.014$~m$^2$/s (best fit, not shown) on that of Eq.~\eqref{fullDispRel}. However, the hollow-core model does not include vertical flow $v_z$, contrary to the (solid-body core) Rankine model~\cite{Thomson1880,Saffman1993} used here. For $m=1$,
Eq.~\eqref{KWeq} reduces to $\omega_{hc}=- \frac{\Gamma}{4\pi}k^2\left[\ln\left(\frac{2}{ka_0}\right)-\gamma \right]$ for $ka_0 \ll 1$, and to $\omega_{hc}=-\frac{\Gamma}{2\pi a_0^{3/2}}k^{1/2}$ for $ka_0 \gg 1$.

\textit{Experimental setup---}Complementary measurements of the azimuthal, $v_{\theta}(r)$, and vertical, $v_z(z)$, velocity fields are performed using particle image velocimetry and particle-tracking velocimetry, respectively, enabling an independent characterization of the background flow and circulation, $\Gamma=2\pi r v_{\theta}(r)$~\cite{BarckickeNaturePhys2026}. This setup closely reproduces the idealized conditions assumed in theoretical descriptions of vortex-line waves, making it a well-controlled laboratory model for Kelvin-wave dynamics~\cite{BarckickeNaturePhys2026}. 
Further details on the experimental setup can be found in our study of linear Kelvin waves~\cite{BarckickeNaturePhys2026}. 



\textit{Experimental and data analysis methods---}The circulation $\Gamma(r) = 2\pi rv_{\theta}(r)$ within the liquid is experimentally obtained from the PIV measurement of $v_{\theta}(r)$ as in Fig.~\ref{Gamma}. The value of $\Gamma$ used throughout the main text is inferred from Fig.~\ref{Gamma} by the intercept of its $r$-linear dependence (dashed line) and the vortex core (gray region), leading to $\Gamma(a_0)=0.018$~m$^2/$s.

\begin{figure}[t!]
\includegraphics[width=0.9\columnwidth]{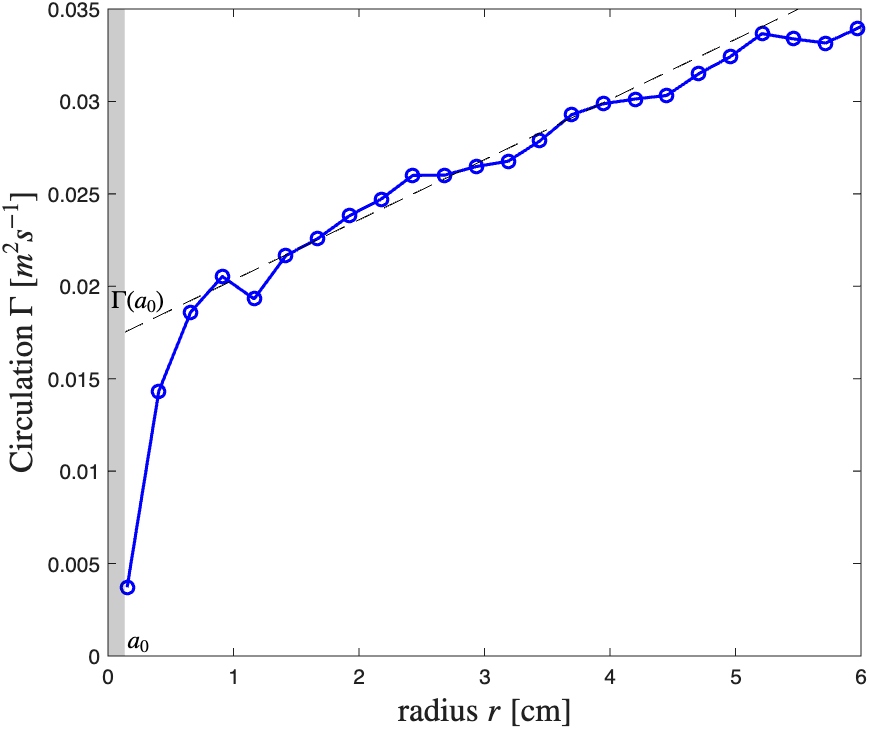} 
\caption{\label{Gamma}Circulation $\Gamma(r) = 2\pi rv_{\theta}(r)$ within the liquid as a function of the radial distance $r$ from the vortex core (in gray). Same pump flow rate as in the main text. $a_0=1.3$~mm. $v_{\theta}(r)$ is measured by PIV at an 18-cm height above the drain. Dashed line: best linear fit leading to $\Gamma(a_0)=0.018$~m$^2/$s.} 
\end{figure}

\begin{figure}[t!]
\includegraphics[width=0.9\columnwidth]{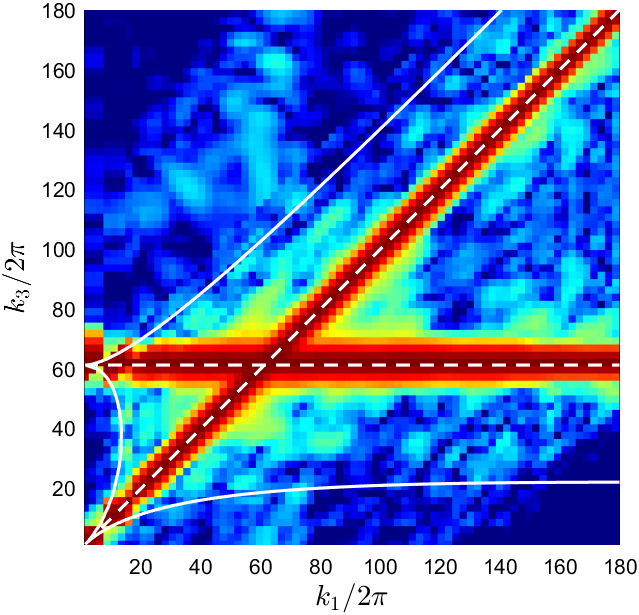} 
\caption{\label{Tri}Tricoherence $T(k_1, k_2, k_3)$ showing no nontrivial four-wave resonant interaction ($k_1+k_2=k_3+k_4$, i.e., $N=4$) for fixed $k_2=60$~m$^{-1}$. Solid lines: resonant interaction solutions of Eq.~\eqref{Nwave} with $N=4$ (with $1\leftrightarrow 3$ in $\omega$, $2\leftrightarrow 2$ in $k$) and $\omega(k)$ as Eq.~\eqref{fullDispRel}. Dashed lines show trivial wave interactions ($k_1=k_3$ and $k_3=k_2$). Logscale colorbar.} 
\end{figure}

\textit{Four-wave resonant interactions---}As Eq.~\eqref{Nwave} has only trivial solutions for $N = 4$ in the case $2\leftrightarrow 2$ in $k$ and $\omega$, we experimentally probe possible exact four-wave resonant interactions, in the case $1\leftrightarrow 3$ in $\omega$ and $2\leftrightarrow 2$ in $k$, as predicted theoretically~\cite{LvovLTP2010}.  To wit, we compute the tricoherence $T(k_1,k_2,k_3)=\frac{|\langle \eta_{k_1}^*\eta_{k_2}^*\eta_{k_3}\eta_{k_1+k_2-k_3}\rangle|}{\sqrt{\langle| \eta_{k_1}\eta_{k_2}|^2\rangle\langle|\eta_{k_3}\eta_{k_1+k_2-k_3}|^2\rangle}}$ as shown in Fig.~\ref{Tri}. Except for trivial solutions (dashed lines), no four-wave interactions occur around the white solid lines, solutions of Eq.~(\ref{Nwave}) with $N=4$ (with $1\leftrightarrow 3$ in $\omega$, $2\leftrightarrow 2$ in $k$) and $\omega(k)$ as in Eq.~\eqref{fullDispRel}.

\textit{Six-wave resonant interactions---}Same qualitative results as in Fig.~\ref{Penta} are found for sextets involving other fixed values of $k_3$, $k_4$ and $k_5$ within the inertial range.


\end{document}